# Computing a Frobenius Coin Problem decision problem in O(n$^2$)

Charles Sauerbier
January 2010

*Abstract:* Expanding on recent results of another an algorithm is presented that provides solution to the Frobenius Coin Problem in worst case O(n$^2$) in the magnitude of the largest denomination.

The Frobenius Coin Problem (FCP) is one of those problems that in the past found little discussion in courses in theoretical computer science or computational complexity, while being somewhat more familiar in studies in mathematics[1]. Recent papers[2] having come into the public arena propel the subject to the fore. The problem has analogs in several other problems[3] more often found discussed in theoretical computer science courses and basic texts on computational complexity.

The Frobenius Coin Problem under certain constraints have been shown by others[4][5] to be of NP-Hard computational complexity. Constraints were not considered by the namesake of the problem, based on the anecdotal evidence[6] of the problems origins. Based on recent results in (Chermakani, 2010), however, the unconstrained FCP decision problem has a worst case complexity of O(n$^2$).

**Unconstrained Frobenius Coin Decision Problem**

Given any set C = { $c_1$, $c_2$, …, $c_n$ } of denominations of coins, represented by $c_i$, what is the largest number for which there does not exist a sum of some subset of the denominations of coins in C.

The problem can be stated mathematically as:

> Given some set C = { $c_1$, $c_2$, …, $c_n$ }, where gdc(C) = 1, and; vector A = < $a_1$, $a_2$, …, $a_n$ >, where $a_i \in \mathbb{Z}$ such that $a_i \geq 0$, and; set S = { s | s ≠ $a_1 c_1 + a_2 c_2 + \ldots + a_n c_n$ $\forall a_i > 0$ }: If S ≠ ∅, what is that value of sup(S) such that sup(S) $\in$ S.

**Chermakani's Theorem**

For the unconstrained FCP we obtain from (Chermakani, 2010) the following as theorem:

> Let C = { $c_1$, $c_2$, …, $c_i$, …, $c_n$ } be some set denominations of coins, where n = |C|. Let L be the largest denomination in C. Let B = < $b_1$, $b_2$, …, $b_j$, …, $b_L$ > be a vector of binary values such that $b_j$ = 1 where j $\in$ C and 0 where j $\notin$ C. Then the number of ways coins of denominations in C can

---

[1] (Beck & Robins, 2007)
[2] (Chermakani, 2010), (Bogart, 2009)
[3] At time of writing http://mathworld.wolfram.com/CoinProblem.html listed a few problems generally viewed as analogous to the Coin Problem.
[4] (Bogart, 2009) cites J.L. Ramirez-Alfonsin "Complexity of the Frobenius problem", *Combinatorica*, 16(1):143-147, March 1996; for which we lack a readily accessible complete copy at time of writing.
[5] (Weisstein) cites R. Kannan, "Lattice Translates of a Polytope and the Frobenius Problem." *Combinatorica* 12:161-177, 1992; for which we lack a readily accessible complete copy at time of writing.
[6] (Bogart, 2009) provides one such instance attributed: J.L. Ramirez-Alfonsin "The Diophantine Frobenius Problem", 2005.

*be arranged in a stack such that the monetary value of the coins in the stack is equal to k is given by the linear recurrence relation:*

$E_k = 0 \; \forall \, k < 0$

$E_0 = 1$

$E_k = b_L E_{k-L} + b_{L-1} E_{k-(L-1)} + \ldots + b_2 E_{k-2} + b_1 E_{k-1} = \sum_{h=1}^{L} b_h E_{k-h} \; \forall \, k > 0.$

Chermakani's theorem is premised on the use of a binary vector of order L representing the elements of B, requiring at least L addition operations. However, in practice the elements in the sum have order $|C| \leq L$, as where $b_j = 0$ for some j the corresponding elements in the computation of $E_i$ containing $b_j$ are zero.

**Frobenius Coin Decision Algorithm[7]**

The consequence of the theorem is that we are provided means to compute the zeros of the Frobenius Number series. The zeros corresponding to an answer to the decision problem where the set S can be established and the value of $\sup(S) \in S$ can be determined. The literature on FCP provides a number of formulae for determining $\sup(S)$ as an upper bound. The formula, as best one can determine, does not assert that the result is in all cases the "Frobenius Number" in the sense that $\sup(S) \in S$.

Given that where $b_j \in B$ is zero the corresponding elements of the summation are also zero, the vector B used in the theorem is not necessary to compute the sum. Direct implementation of the theorem as an algorithm is presented in *Algorithm 1* (below). This naïve implementation of the theorem as an algorithm, on inspection appears to be of polynomial time complexity with a worst case $O(n^2)$. Yet, it fails to be polynomial time–space due to requiring exponential space and consequently exponential time in execution due to the growth of the value that is the result of summation.

```
Algorithm 1
Input:
1. A set C={c1, c2, …, cn}
2. F - upper bound on Frobenius Number

Let C = {c1, c2, …, cn}
Let E = {1}
Let L = MAX(C)
Let N = |C|
Let y = 0
Loop 1 {  ⎧ For i = 1 To F
          |    x = 0
          |           ⎧ For j = 1 To N
          |  Loop 2 { |    x += E[i-C[j]]
          |           ⎩ Next
          |    E[i] = x
          |    If (0 == x)
          |       y = i
          |    End If
          ⎩ Next
If (0 < y)
   Output y
End If
```

The exponential time–space problem can be addressed where our interest is only in determining the existence of zeros and the values at which the zeros occur. Such is all that is necessary and sufficient to the decision problem. Limiting the objective of the computation to such determination allows a minor refinement that eliminates the problem of exponential growth in the computed value, and thus in the space of the problem. *Algorithm 2* (below) presents the algorithm so modified.

---

[7] One can check the results of Chermakani's Theorem using a spreadsheet.

Where the naïve implementation of *Algorithm 1* requires exponentially increasing space to retain the computed values and thus time to compute the sum at each pass of *Loop 1*, *Algorithm 2* reduces the execution space requirements to a linearly increasing value and time required to perform the sum to a constant. *Algorithm 3* (below) further refines the solution by reducing the execution space complexity to a constant, while increasing the computational effort by a negligible amount. The change from *Algorithm 2* is to restrict the dimension of the array of computed values to L + 1 and shift downward the computed values; thus retaining only those necessary to the next iteration. This change leaves the space complexity of the decision problem linear in the magnitude of L, while holding the execution space complexity of the algorithm to a fixed constant value.

```
Algorithm 2
Input:
1. A set C={c1, c2, …, cn}
2. F - upper bound on Frobenius Number

Let C = {c1, c2, …, cn}
Let E = {1}
Let L = MAX(C)
Let N = |C|

Let Y = 0
        ⎡ For i = 1 To F
        |    x = 0
        |         ⎡ For j = 1 To N
        | Loop 2 ⎨    x += E[i-C[j]]
        |         ⎣ Next
Loop 1 ⎨    E[i] = x
        |    If (0 == x)
        |        E[i] = 0
        |        y = i
        |    Else
        |        E[i] = 1
        |    End If
        ⎣ Next
If (0 < y)
    Output y
End If
```

```
Input:
1. A set C={c₁, c₂, …, cₙ} representing
   denominations of coins such that
   gcd(C) = 1.
2. F the upper bound on Frobenius Number
   computed by any of several methods
   available in literature.

Let L = MAX(C) be the element having
         the maximum value in the set C.
Let N = |C| be the number of elements in C.
Let C = {c₁, c₂, …, cₙ} be an array
         containing input C.
Let E = {1, 0, 0 …} be an array of
         size L + 1 of computed values of E.
```

```
Algorithm 3

Let y = 0
Let z = L + 1
        ⎡ For i = 1 To F
        |    x = 0
        |         ⎡ For j = 1 To N
        | Loop 2 ⎨    x += E[z-C[j]]
        |         ⎣ Next
Loop 1 ⎨    If (0 == x)
        |        E[i] = 0
        |        y = i
        |    Else
        |        E[i] = 1
        |    End If
        |         ⎡ For k = 2 To z
        | Loop 3 ⎨    E[k-1] = E[k]
        |         ⎣ Next
        ⎣ Next
If (0 < y)
    Output y
End If
```

## A Dichotomy of Complexity

The result here is that we can now directly compute the Frobenius Number, as where $F_i$ = 0 the value of $E_i$ = 0 and conversely so, by means in P time–space complexity. Further, given any upper limit X on a sum of coin denominations we can determine the decision problem of whether there exists a sum of coins of denominations in C that is equal to X by determining whether X is a zero point in the sequence computed by Chermakani's Theorem.

That result is interesting in that determining a solution of the unconstrained FCP as a decision problem for the question of whether $\sup(S) \in S$ exists, the complexity of the FCP functional problem of determining a instance of the vector $A = <a_1, a_2, …, a_n>$ is not resolved. The nature of the solution presented does not readily provide means to determine values for elements of A. The solution also does not resolve or lend itself to resolution of the FCP counting problem (i.e.: determining the number of possible solutions) as the solution then reverts to being exponential space in execution.

The algorithm here also does not provide means to readily resolve any of several constrained versions of the Frobenius Coin Problem, nor does it provide means to solve the "Coin Exchange Problem" (CEP). The CEP problem extends allowed values for elements of A to include negative numbers.

That sup(S) exists in the absence of constraint on the number of coins of each denomination is known by consequence of determining that gcd(C) = 1. A question resolved by computing the $\gcd(c_i, c_j)$ for all i, j; a problem with a known polynomial time–space solution. The existence of means to compute a value for $\sup(S) \in S$, given best available evidence in hand to date, lacks a P time–space algorithm for determination of either.

**Conclusion**

Recent results presented in (Chermakani, 2010) provide basis for derivation of algorithmic means to determine the Frobenius Number for any given set of denominations of coins with an execution complexity that is polynomial in time–space with worst cast $O(n^2)$ in the magnitude of the upper bound on F as determined by means available in literature[8] shown to be linear in the magnitude of maximum denomination of coins. The results, while solving the decision problem by determining a value for the unconstrained Frobenius Coin Problem, do not answer the question of P = NP as the function problem of determining an assignment of coins is unresolved; as is the FCP counting problem and constrained variations of FCP the solution presented here does not resolve.

---

[8] See references in (Acosta)

**Appendix**

**A. Algorithms**

**A.1 Algorithm 1**

```
Algorithm 1
Input:
1. A set C={c1, c2, …, cn}
2. F - upper bound on Frobenius Number

Let C = {c1, c2, …, cn}
Let E = {1}
Let L = MAX(C)
Let N = |C|
Let y = 0
          ⎡  For i = 1 To F
          |      x = 0
          |          ⎡  For j = 1 To N
          |   Loop 2 {     x += E[i-C[j]]
          |          ⎣  Next
Loop 1  { E[i] = x
          |      If (0 == x)
          |          y = i
          |      End If
          ⎣  Next
If (0 < y)
   Output y
End If
```

## A.2 Algorithm 2

```
Algorithm 2
Input:
1. A set C={c1, c2, …, cn}
2. F - upper bound on Frobenius Number

Let C = {c1, c2, …, cn}
Let E = {1}
Let L = MAX(C)
Let N = |C|

Let Y = 0
            ⎡  For i = 1 To F
            |     x = 0
            |           ⎡  For j = 1 To N
            |  Loop 2 {      x += E[i-C[j]]
            |           ⎣  Next
Loop 1 {       E[i] = x
            |     If (0 == x)
            |        E[i] = 0
            |        y = i
            |     Else
            |        E[i] = 1
            |     End If
            ⎣  Next
If (0 < y)
   Output y
End If
```

## A.3 Algorithm 3

```
Input:
1. A set C={c₁, c₂, …, cₙ} representing
   denominations of coins such that
   gcd(C) = 1.
2. F the upper bound on Frobenius Number
   computed by any of several methods
   available in literature.

Let L = MAX(C) be the element having
        the maximum value in the set C.
Let N = |C| be the number of elements in C.
Let C = {c₁, c₂, …, cₙ} be an array
        containing input C.
Let E = {1, 0, 0 …} be an array of
        size L + 1 of computed values of E.
```

```
Algorithm 3

Let y = 0
Let z = L + 1
         ⎡ For i = 1 To F
         |   x = 0
         |            ⎡ For j = 1 To N
         |   Loop 2 {   x += E[z-C[j]]
         |            ⎣ Next
Loop 1 {    If (0 == x)
         |       E[i] = 0
         |       y = i
         |    Else
         |       E[i] = 1
         |    End If
         |            ⎡ For k = 2 To z
         |   Loop 3 {   E[k-1] = E[k]
         |            ⎣ Next
         ⎣ Next
If (0 < y)
   Output y
End If
```